\documentclass[conference]{IEEEtran}

\ifCLASSINFOpdf
\else
\fi

\usepackage{times}
\usepackage{helvet}
\usepackage{courier}
\usepackage{booktabs}
\usepackage{graphicx}
\usepackage{makeidx}  % allows for indexgeneration
\usepackage{subfig}
\usepackage{amssymb,amsmath,amsfonts}
\usepackage{url}

\begin{document}
%
% paper title
% Titles are generally capitalized except for words such as a, an, and, as,
% at, but, by, for, in, nor, of, on, or, the, to and up, which are usually
% not capitalized unless they are the first or last word of the title.
% Linebreaks \\ can be used within to get better formatting as desired.
% Do not put math or special symbols in the title.

\title{On Analyzing Job Hop Behavior and Talent Flow Networks}
%\author{AAAI Press\\
%Association for the Advancement of Artificial Intelligence\\
%2275 East Bayshore Road, Suite 160\\
%Palo Alto, California 94303\\
%}

% author names and affiliations
% use a multiple column layout for up to three different
% affiliations
\author{}

\author{\IEEEauthorblockN{Richard J. Oentaryo\thanks{The work was carried out during Richard's stay in Singapore Management University.}}
\IEEEauthorblockA{McLaren Applied Technologies, Singapore\\
Email: richard.oentaryo@mclaren.com}
\and
\IEEEauthorblockN{Xavier Jayaraj Siddarth Ashok, Ee-Peng Lim, Philips Kokoh Prasetyo}
\IEEEauthorblockA{Singapore Management University\\
Email: \{xaviera,eplim,pprasetyo\}@smu.edu.sg}}

%\and
%\IEEEauthorblockN{Firstname Lastname\\ and Firstname Lastname}
%\IEEEauthorblockA{School Affiliation\\
%University Affiliation\\
%State, Country}}

% conference papers do not typically use \thanks and this command
% is locked out in conference mode. If really needed, such as for
% the acknowledgment of grants, issue a \IEEEoverridecommandlockouts
% after \documentclass

% for over three affiliations, or if they all won't fit within the width
% of the page, use this alternative format:
%
%\author{\IEEEauthorblockN{Michael Shell\IEEEauthorrefmark{1},
%Homer Simpson\IEEEauthorrefmark{2},
%James Kirk\IEEEauthorrefmark{3},
%Montgomery Scott\IEEEauthorrefmark{3} and
%Eldon Tyrell\IEEEauthorrefmark{4}}
%\IEEEauthorblockA{\IEEEauthorrefmark{1}School of Electrical and Computer Engineering\\
%Georgia Institute of Technology,
%Atlanta, Georgia 30332--0250\\ Email: see http://www.michaelshell.org/contact.html}
%\IEEEauthorblockA{\IEEEauthorrefmark{2}Twentieth Century Fox, Springfield, USA\\
%Email: homer@thesimpsons.com}
%\IEEEauthorblockA{\IEEEauthorrefmark{3}Starfleet Academy, San Francisco, California 96678-2391\\
%Telephone: (800) 555--1212, Fax: (888) 555--1212}
%\IEEEauthorblockA{\IEEEauthorrefmark{4}Tyrell Inc., 123 Replicant Street, Los Angeles, California 90210--4321}}

% use for special paper notices
%\IEEEspecialpapernotice{(Invited Paper)}

% make the title area
\maketitle

\begin{abstract}
Analyzing job hopping behavior is important for the understanding of job preference and career progression of working individuals.  When analyzed at the workforce population level, job hop analysis helps to gain insights of talent flow and organization competition.  Traditionally, surveys are conducted on job seekers and employers to study job behavior. While surveys are good at getting direct user input to specially designed questions, they are often not scalable and timely enough to cope with fast-changing job landscape. In this paper, we present a data science approach to analyze job hops performed by about 490,000 working professionals located in a city using their publicly shared profiles. We develop several metrics to measure how much work experience is needed to take up a job and how recent/established the job is, and then examine how these metrics correlate with the propensity of hopping. We also study how job hop behavior is related to job promotion/demotion. Finally, we perform network analyses at the job and organization levels in order to derive insights on talent flow as well as job and organizational competitiveness.
%\keywords{Data analytics, job hop, network analysis, social media}
\end{abstract}

% no keywords

% For peer review papers, you can put extra information on the cover
% page as needed:
% \ifCLASSOPTIONpeerreview
% \begin{center} \bfseries EDICS Category: 3-BBND \end{center}
% \fi
%
% For peerreview papers, this IEEEtran command inserts a page break and
% creates the second title. It will be ignored for other modes.
\IEEEpeerreviewmaketitle

\section{Introduction}
\label{sec:introduction}

%\textbf{Motivation.}
%Workforce employability is a key concern of many governments today, and substantial amount of efforts and resources have been spent to boost job creation and skill training. The success of such efforts depends on how much insights one has on the availability of jobs and skills as well as the career movement of workforce, before a suitable plan is developed.
Job hop is a common behavior observed in any workforce. As a person hops from jobs to jobs, he or she acquires new skills and potentially gains higher income.  Every job hop captures an important decision made by the person as well as an attempt of the hiring organization to acquire talent. When job hop behavior is analyzed at the workforce level, it will yield insights about the workforce, job pool and employers.

Such insights have been traditionally obtained using surveys on employers and job seekers. For example, the US Bureau of Labor Statistics (BLS) conducts annual surveys with approximately 146,000 businesses and government agencies to collect employment data\footnote{\scriptsize \url{www.bls.gov/ces/}}.  The surveys yield useful information about job demand, job supply, income, working hours, etc..  While surveys can be a powerful instrument to gather direct user input, they are usually not scalable.  In the case of the BLS surveys, they cover less than $1\%$ of all U.S businesses. Moreover, as fast-changing technologies (such as sharing economy \cite{Hamari2015}) begin to impact job demand quickly, it is critical to explore new ways to obtain job related insights.

Past studies \cite{Joseph:2012,Maier:2015} also tend to study jobs and organizations as isolated entities, without considering them as connected networks which capture talent flows from jobs to jobs and from organizations to other organizations. A lack of this network view prevents us from analyzing the ways people build their career, and competition among organizations for talent. For example, some job changes could be promotions, while others could just be lateral and even demotions. The network view is also crucial in studying the competitions among jobs and organizations that eventually impact job creation and talent attraction.

In contrast, online professional networks (OPNs) are fast becoming a marketplace for resume posting, candidate hunting, and job searching. Representative examples of OPN are LinkedIn, Xing and Viadeo\footnote{\scriptsize \textbf{LinkedIn} -- \url{www.linkedin.com}; \textbf{Xing} -- \url{www.xing.com}; \textbf{Viadeo} -- \url{www.viadeo.com}}. A lot of detailed job activity data at the individual user level are now publicly available in the OPNs, as soon as the users update their profiles. These data can be analyzed to derive interesting behavioral insights about jobs and organizations, as well as to build services that can benefit both employers and job seekers, e.g., a service that helps employers find suitable employees and job seekers find suitable jobs.

% RICHARD: Sorry Prof, I guess we better remove the paragraph below, as we do not want to specifically mention LinkedIn, and also because this claim is inaccurate - LinkedIn do publish their findings about economic graph and talent flow. See our related works section.
%
% LinkedIn is a very popular social media platform that attracts more than $100$ millions users worldwide.  It offers services to both employers and job seekers, helping employers find suitable employees and job seekers find  suitable jobs. However, LinkedIn does not offer public insights on career movement patterns that can increase the awareness of workforce movement, talent flow, and organization level competition for skilled workforce. Despite the importance of workforce to today's economy, there is very little analysis on their workforce movement and its impact to talent management and organization competition.

%end revision

\textbf{Objectives}.
In this work, we therefore focus on using data from one of the world's largest OPNs to analyze job hops and talent flow. To support our analyses on hops within an organization and those involving different organizations, we first develop several metrics that measure the amount of experience is required for every job and its age, from the perspective of people holding the job.
% In this research, we therefore focus on analyzing jobs, workforce and career movement of workforce at a large scale using public profiles of LinkedIn users. In our analysis, we aim to characterize the jobs held by users from a city state.  This gives us a sense of \richard{how new and how good each job is} with respect to the workforce population. This analysis is important to job creation and skill training policy making and implementation.

We also aim at studying how the job hop behavior of a workforce is related to job promotion/demotion. This is a topic often discussed based on anecdotal examples \cite{Alper:1994,Hamori:2010}. A better approach is to conduct a large-scale data analytics study. This will give much broader insights on job hop patterns particularly useful in human resource recruitment and career coaching. %In this work, we focus on only the LinkedIn users in Singapore.

Finally, our research aims at analyzing talent flow based on job hop behavior and measuring the capabilities of each job and organization in attracting, supplying, and competing for human capital. To this end, we create a weighted directed hop network among jobs and organizations, develop different centrality measures for the job and organization nodes, and evaluate them by manual inspection or by comparing with other attributes such as organization size.

% RICHARD: Sorry Prof, we should probably remove the paragraph below as well.
% \textbf{Challenges}.
% Accomplishing the above three objectives is challenging for several reasons.  Firstly, one has to obtain the necessary user profiles of the entire Singapore workforce.  This requires systematic data crawling and data preprocessing.  Data preprocessing is required to remove data noises (e.g., incorrect company names, incorrect job titles, etc.), to extract only the required data from LinkedIn while repurposing the extracted data for job characterization, and job hop behavior analysis.  For evaluation purposes, ground truth labels normally do not exist in the data itself.  We thus have to infer some attributes by other external means, similar to many other data science research.

\textbf{Contributions}.
We summarize our key contributions as follows:
\begin{itemize}
\item We present a new empirical study on job hops involving a city-scale workforce sharing data on an OPN. Unlike past survey works \cite{NBERw2649,LabourMobility,moscarini2007occupational,fuller2008job,Joseph:2012,Schawbel:2013} and more recent data-oriented studies \cite{Xu:KDD2016,Kapur:KDD2016,Liu:AAAI2016,Chaudhury:WWW2016,Xu:ICDM2015}, our work offers broader analysis that is not constrained to specific workforce segments or industries.

\item We develop a data analytics methodology for analyzing job hops, in which we measure the work experience requirement and recency/establishment of a job as well as how they relate to the propensity of hopping. Based on these measures, we also quantify the level gain of a hop, which allows us to analyze job promotion/demotion in relation to hops within and across organizations.

\item We analyze talent flow across jobs and companies by constructing job-level and organization-level hop networks respectively. We develop and evaluate several centrality metrics that measure the extent to which jobs and organizations attract, supply and compete for human capital.
\end{itemize}

\textbf{Paper outline}.
In Section \ref{sec:related_work}, we first review related works on job analysis. Section \ref{sec:approach} presents our data analytics methodology. We then elaborate our empirical results and findings in Section \ref{sec:results}. Finally, we conclude in Section \ref{sec:conclusion}.

\section{Related Work}
\label{sec:related_work}

Research on job and workforce movements has been around for decades \cite{NBERw2649,LabourMobility,moscarini2007occupational,fuller2008job,Joseph:2012}.  Topel \emph{et al.} \cite{NBERw2649} analyzed $15$ years of job changing and wage growth of young men from Longitudinal Employee-Employer data. Long \emph{et al.}~\cite{LabourMobility} studied the labor mobility in Europe and the U.S. Moscarini \emph{et al.} \cite{moscarini2007occupational} measured worker mobility across occupations and jobs in the monthly Current Population Survey data from 1979 to 2006. More recent survey-based studies \cite{fuller2008job,Joseph:2012,Schawbel:2013} have revealed that the younger employees are more likely to switch jobs and employers/companies than the older ones.
%Fuller~\cite{fuller2008job} and Joseph \emph{et al.} examined the National Longitudinal Survey of Youth data from 1979 to 2002. 
All these studies traditionally relied on surveys, census, and other data such as tax lists and population registers, which require extensive and time-consuming efforts to collect. Moreover, the findings are usually biased to selected workforce segments or industries, and cannot be easily scaled up or replicated in other segments/industries.

% Joseph \emph{et al.} studied the job trajectories of $500$ individuals and found that the older employees are less likely to change jobs and the organizations they work for than the younger ones \cite{Joseph:2012}.  Other surveys on selected workforce segments revealed that younger employees are more prone to job hops to other companies \cite{Schawbel:2013}.  
% The past studies of job hops have been often based on a selected industry (e.g., university faculty \cite{Alper:1994}, employees in financial industry \cite{Hamori:2010}, IT professionals \cite{Joseph:2012}). 

With the wide adoption of OPNs, there is a rapidly-growing interest to mine the online user data from the OPNs to understand job and workforce movements as well as career growth. For example, State \emph{et al.} \cite{State:Socinfo2014} analyzed the migration trends of professional workers into the U.S. Xu \emph{et al.} \cite{Xu:ICDM2015} combined work experiences from OPNs and check-in records from location-based social networks to predict job change occasions. Chaudhury \emph{et al.} \cite{Chaudhury:WWW2016} analyzed the growth patterns of the ego-network of new employees in companies.

An important aspect in OPNs is job hop. Job hop data capture a wide range of signals that can help understand the performances of organizations, talent sources, job market, professional profiles, as well as career advancement. Cheng \emph{et al.} \cite{Cheng:KDD2013} modeled job hop activities to rank influential companies. Xu \emph{et al.} \cite{Xu:KDD2016} generated and analyzed job hop networks to identify talent circles. Kapur \emph{et al.} \cite{Kapur:KDD2016} devised the Talent Flow Graph to rank universities based on the career outcomes of their graduates. They applied their approach to two specific workforce segments: investment banker and software developer. %The ranking is achieved based on an adaptation of the PageRank algorithm \cite{Brin1998}.

Users' career paths have also been utilized to model professional similarity for use in job recruitment process \cite{Xu:KDD2014}. In this work, a sequence alignment method was used to quantify similarity between two career paths. Liu \emph{et al.} \cite{Liu:AAAI2016} devised a multi-source learning framework that combines information from multiple social networks to predict the career path of a user. Their work focused on four job categories: software engineer, sales, consultant, and marketing. 
%In a recent study, LinkedIn also analyzed job hop activities in the U.S, and reported that \richard{job hopping has dramatically increased over the years, that hopping is more common in certain industries, and that women hop more than men\footnote{\scriptsize \url{https://business.linkedin.com/talent-solutions/blog/trends-and-research/2016/job-hopping-has-increased--and-will-accelerate}}}.

\textbf{Our research}. The work presented in this paper differs from the above-mentioned works in several unique ways. Firstly, we introduce quantitative metrics to measure how much work experience is required to take up a job and how recent/established a job is, and examine their relationships with the propensity of hopping. Secondly, we compute the level gain of job hops so as to analyze promotion/demotion of employees which, to our best knowledge, has been missing in the previous studies. Finally, we perform an extensive study on talent flow and competition by analyzing both job-level and organization-level hop networks, without restricting ourselves to specific workforce segments or industries.
\section{Analytics Methodology}
\label{sec:approach}

%This section describes our methodology for analyzing job hops using OPN data.
%We begin by presenting our data harvesting method and basic terminologies. We then present our hop graph construction method as well as key job metrics.

\subsection{Data Harvesting}
\label{sec:data_harvesting}

In this work, we study the job hop data extracted from one of the largest OPNs.  Such data are not generally available and technically challenging to gather.  To give a meaningful scope to our study within some resource constraint, we decide to cover all public profiles of the OPN users located in a target city within the First World economy.  The data was collected around $30$ June $2016$. Specifically, the data consist of (1) all public user profiles that found in the directory of users associated with the target city, %(e.g., https://sg.linkedin.com/directory/people-a/)
and (2) organization profiles that are mentioned in these public user profiles.
%(cf. https://sg.linkedin.com/directory/organizations/).
In our study, we focus on \emph{active profiles}, defined as user profiles with least one entry in the \emph{education} and \emph{skills} fields. Table~\ref{tab:basic_stats} summarizes the data statistics.

It is worth noting that, while our dataset covers a comprehensive set of user profiles, it may still suffer from population bias \cite{Olteanu2016}. That is, the data does not necessarily capture all sorts of occupations. Our dataset, for instance, may leave out blue-collar/non-technical workers who do not use social media. Nevertheless, our OPN data are arguably representative of all professionals, managers, executives and technicians (PMET), who increasingly make up the majority of working population in a developed city economy.
%\footnote{\scriptsize \url{http://www.nptd.gov.sg/portals/0/news/population-white-paper.pdf}}.

\begin{table}[!t]
%\scriptsize
\centering
\caption{Dataset Statistics}
\label{tab:basic_stats}
\begin{tabular}{|l|c|c|}
\hline
Statistics & Value \\
\hline
No. of user profiles 		& 2,574,502\\
No. of active user profiles & 490,200\\
No. of organizations   			& 145,524\\
No. of industries         	& 147\\
\hline
\end{tabular}
\end{table}

To facilitate data collection, we devise a data crawler that performs two steps to collect the URLs of public user profiles and organization profiles from the OPN website. Firstly, we collected all user profile URLs from the user directory of the target city.  We then crawled the content of each user profile using the collected profile URLs.
%It focuses on gathering such information as name, title, number of connections, summary, skills, work experience, education, projects involved, courses, certifications, languages, interests, publications, groups and organizations the user interested in, honours and volunteering work from public user profiles.
Note that our crawler does not capture users' online friends nor wall posts. Finally, while crawling the user profiles, the organization URLs found in each user profile are collected and used to crawl the organization profiles (pages).
%Information such as name, overview, description, specialities, parent company and other brand pages are extracted when crawling the company profiles.

\subsection{Notations and Definitions}
\label{sec:hop_definition}

We derive job hops from job history in the user profiles.  We first denote a \emph{job} as a tuple $(t, c, i)$, which means a job title $t$ at organization $c$ in industry $i$. Note that each organization $c$ belongs to a unique industry $i$. We then define a \emph{hop} as a transition from one job to another with \emph{non-overlapping} time period.
%Figure~\ref{fig:hop_definition} shows an example of user who lists five jobs A, B, C, D and E in his profile.
%Suppose a user is regarded as having only three hops, i.e., from job A to job B, from B to E, and from C to D. There is no hop from B to C, or from B to D, or from D to E, as these are jobs from different parallel tracks of job activities. %We also note that a user may have multiple \emph{latest jobs}, i.e., jobs D and E in this example.

Based on the above, we further distinguish between two types of hop:
\begin{itemize}
\item \textbf{External hop}. This is defined as a move from one job to another job, where the source and destination organizations are \emph{different}. Formally, an external hop is a hop from job $(t,c,i)$ to job $(t',c',i')$ where $c \neq c'$. Here the source job title $t$ can be either the same as or different from the destination job title $t'$.

\item \textbf{Internal hop}. This refers to a move from one job to another, where the source and destination organizations are \emph{the same}. That is, an internal hop is a hop from job $(t,c,i)$ to job $(t',c',i')$ where $c = c'$. To avoid duplicates (e.g., a person may state three times that (s)he is a Civil Engineer at organization X, as (s)he has worked on three construction projects under the same organization), however, we add a constraint $t \neq t'$. As such, a move from $(t,c,i)$ to $(t',c',i')$ where $t = t'$ and $c = c'$ is \emph{not} counted as a (valid) internal hop.
\end{itemize}

%\begin{figure}[!t]
%\centering
%\includegraphics[width=0.3\textwidth]{figures/hop_definition.pdf}
%\caption{Our definition of job hop}
%\label{fig:hop_definition}
%\end{figure}

\subsection{Hop Graph Construction}
\label{sec:hop_network}

After establishing the job hops for all user profiles in our OPN data, we construct two types of \emph{weighted directed graphs} to facilitate our analyses, namely: (1) \textbf{job hop graph} and (2) \textbf{organization hop graph}. Each node $v_{t,i}$ in the job hop graph (or simply, job graph) represents a job title $t$ in industry $i$, while a node $v_{c}$ in the organization hop graph (or simply, organization graph) refers to a organization $c$.

For the job graph, a directed edge is created from node $v_{t,i}$ to node $v_{t',i'}$ if there is at least one person moving from a (job title,industry) pair $(t,i)$ to another pair $(t',i')$. We also capture the number of user profiles moving from $(t,i)$ to $(t',i')$ as the \textbf{edge weight} $e_{(t,i) \rightarrow (t',i')}$. The same applies to the organization graph, i.e., the edge weight $e_{c \rightarrow c'}$ represents the number of users moving from an organization $c$ to another organization $c'$.

Finally, to handle noise due to data sparsity (e.g., unusual/spurious job titles or organizations), we define a \textbf{minimum support} threshold for each node in the job graph and organization graph. Nodes with the number of users less than the minimum support will be removed from both graphs. Unless explicitly specified, we shall use the default minimum support of $10$ users in our empirical study.

\subsection{Key Metrics}
\label{sec:metrics}

In our study, we want to tell how much career advancement people make in their jobs.  We therefore need to first estimate the experience of a person holding a job. Secondly, to determine changes of job market over time, we need to estimate how long a job has existed.
To fulfill the two goals, we introduce the following key metrics respectively, which are applied to active user profiles (as defined in Section~\ref{sec:data_harvesting}):
\begin{itemize}
  \item \textbf{Work experience}: This refers to the duration since the last graduation date of a person till the time at which (s)he finishes a particular job. For a person $p$ with job title $t$ at organization $c$, the work experience is:
    \begin{align}
    wk\_exp(p, t, c, i) = end\_time(p, t, c, i) -
    grad\_date(p)
    \end{align}
    where $grad\_date(p)$ denotes the last graduation date mentioned in his profile.

    For a given job title $t$ from industry $i$, the work experience of the (job title, industry) pair $(t,i)$ is therefore:
    \begin{align}
    wk\_exp(t, i) = Avg_{(p,c)} wk\_exp(p,t,c,i)
    \end{align}

    Examples of job with high $wk\_exp$ score according to our data are ``Professor'', ``Managing Director'', and ``CEO'', whereas examples of job with low $wk\_exp$ score are ``Intern'' and ``Teaching Assistant''.

  \item \textbf{Job age}: This is the duration from the start of a given job until the current date $curr\_date$. It measures how recent or established a job is from the perspective of a person holding the job. For a person $p$ with job title $t$ at organization $c$, the job age is defined as:
    \begin{align}
    job\_age(p, t, c, i) = curr\_date - start\_date(p, t, c, i)
    \end{align}
    where $start\_date(p,t,c,i)$ refers to the start date of the person $p$'s job title $t$ at organization $c$ of industry $i$.
    For a given job title $t$ from industry $i$, the age of the (job title, industry) pair $(t,i)$ is therefore:
    \begin{align}
    job\_age(t, i) = Avg_{(p,c)} job\_age(p,t,c,i)
    \end{align}
    Examples of job with high $job\_age$ score based on our data are ``Director'', ``Systems Engineer'', and ``Division Manager'', whereas examples of job with low $job\_age$ are ``Data Scientist'' and ``Media Analyst''.

% \item \textbf{Skill count}: The number of skills listed in a user's profile. These skills are either specified by the user him/herself or endorsed by his/her connections. For a person $p$, we denote the skill count as $skill\_count(p)$.

\end{itemize}

Based on the above metrics, we further derive several higher-level metrics by aggregating over user profiles at either the job or organization level, as follows:
\begin{itemize}
\item \textbf{External hop fraction}: The fraction of people who move out from a organization $c$ to a different organization $c' \neq c$ over the (total) people hopping from organization $c$. Formally, for a given group of users $g$ (e.g., work experience, job age, or skill count group), the external hop fraction is:
\begin{align}
\label{eqn:ext_hop_frac}
\%external\_hop(g) = \frac{ | \mathbf{P}_{c \rightarrow c'}^g | }{ | \mathbf{P}_{c \rightarrow c'}^g | + | \mathbf{P}_{c \rightarrow c}^g | }
\end{align}
where $\mathbf{P}_{c \rightarrow c'}^g$ is the set of all user profiles belonging to group $g$ who perform \emph{external hops} from some arbitrary organizations $c$ to \emph{different} organizations $c' \neq c$. Conversely, $\mathbf{P}_{c \rightarrow c}^g$ is the set of user profiles belonging to group $g$ who perform \emph{internal hop} within the \emph{same} organization $c$.

\item \textbf{Job level}: As different organizations offer jobs of different rewards and seniority levels (even for the same job titles), we want to be able to measure them.  Since our data do not carry any salary information, we estimate the seniority level of a job $(t,c)$ by computing the \emph{average work experience} over all users who mention job title $t$ at organization $c$ in their profiles as follows:
    \begin{align}
    job\_level(t,c) &= \frac{1}{|\mathbf{P}_{t,c}|} \sum_{p \in \mathbf{P}_{t,c}} wk\_exp(p,t,c,i)	\end{align}
    where $\mathbf{P}_{t,c}$ is the set of all people who include job $(t,c)$ in their profiles. In the equation, $i$ can be inferred from $c$.  Intuitively, a job with longer average work experience implies that a longer time is required to achieve that position, and hence we can expect it to be a high-level job (e.g., CEO of a multi-national organization).

\item \textbf{Level gain}: This refers to the difference between the levels of two jobs within the same or different companies. A positive level gain can be loosely interpreted as a ``promotion'', whereas a negative level gain loosely implies a ``demotion''. Here the ``promotion'' (``demotion'') does not necessarily mean a monetary increase (decrease), but more of an increase (decrease) in the level of work experience required. Formally, the level gain for hop from job $(t,c)$ to job $(t',c')$ is given by:
    \begin{align}
    level\_gain((t,c), (t',c')) = job\_level(t',c') - \nonumber \\
    job\_level(t,c)
    \end{align}
    We note that, although there is no ground truth available in our OPN data, our manual inspections show that the level gain provides a reasonable proxy for a promotion or demotion. It is also worth mentioning that we do not find zero level gain (i.e., neither ``promotion'' or ``demotion'') in our data.
\end{itemize}

We next introduce network centrality metrics to measure node importance in both job and organization graphs.
\begin{itemize}
\item \textbf{In-degree centrality}. This metric refers to the number of inbound (unweighted) edges for a node in the job or organization graph. The in-degree centrality can be interpreted as a measure of how \emph{prominent} a job (or organization) is in a local sense---a high in-degree may imply that it attract talents from the immediate in-neighbors. For this metric, we do not take into account the edge weight information (i.e., the total number of incoming user profiles), as we want to minimize the support bias due to a large number of users for a given job (organization).

\item \textbf{Out-degree centrality}. This is defined as the number of outbound (unweighted) edges for a node in the job or organization graph. We can use the out-degree centrality to measure how \emph{influential} a job (or organization) is in a local sense---a high out-degree may be indicative of a talent supplier to the immediate out-neighbors. Again, we do not utilize the edge weight to compute this metric, so as to mitigate the support bias.

\item \textbf{PageRank centrality}. This is a well-known metric originally used to rank web pages \cite{Brin1998}.
    %, but now also used in wider applications such as social network analysis and recommendation system \cite{Gleich2015}.
    PageRank views inbound edges as ``votes'', and the key idea is that ``votes'' from important nodes should carry more weight than ``votes'' from less important nodes.
    %Also, the significance of a ``vote'' from any source node should be scaled by the number of outbound edges the source node is linking to.
    In this work, we employ a \emph{weighted} version of PageRank \cite{Langville2005}, whereby the transition probabilities for each (source) node is proportional to the (out-)edge weights divided by the weighted out-degree of the node. In the context of job and organization graphs, the weighted PageRank can be viewed as a measure of \emph{global competitiveness}---a job or organization with high PageRank reflects a ``desirable'' destination point where the flow of talent is heading to. In this case, we use edge weight as the hop volume matters in determining where the flow goes to. To avoid dead ends (i.e., nodes with zero out-degree), we also allow our PageRank to perform random jump with the default ``teleportation'' probability of $0.15$.
\end{itemize}

\section{Results and Insights}
\label{sec:results}

In this section, we present our empirical findings and analyses, based on the methodology and metrics described in Section~\ref{sec:approach}.

\subsection{Distribution Analysis}

We first examine the distributions of several basic metrics, including skill count, work experience, and job age.  We found that the active user profiles most commonly have $10$--$15$ skills. The anomaly at $50$ skills is due to the fact that our OPN imposes a maximum limit of $50$ skills per user profile. Secondly, most jobs consist of young workforce, who have work experience of $5$ years or less.  Most users in our OPN data are relatively young in terms of work experience.  These could be due to the younger users showing more interest in using OPN to conduct professional networking.  On the other hand, there are only very few people who have worked for over $20$ years. Here the most common work experience (i.e., the mode) is $2$ years.

Finally, most jobs have been established for $1$ year or more. On the other hand, only very few jobs have been established for more than $20$ years. As with work experience, the most common job age is $2$ years.  The relatively young job age can be explained partly by the young user base, and partly by the sparsity of old but senior-level jobs.  From the labor economics perspective, the findings also suggest attention to be given to identifying and creating more senior jobs to support an ageing workforce.
%
% \begin{figure}[!t]
% \centering
% \includegraphics[width=1.0\textwidth]{figures/hop_fraction_global.pdf}
% \caption{Hop fraction distributions}
% \label{fig:hop_frac_dist}
% \end{figure}

\begin{figure*}[!t]
\centering
\includegraphics[width=0.85\textwidth]{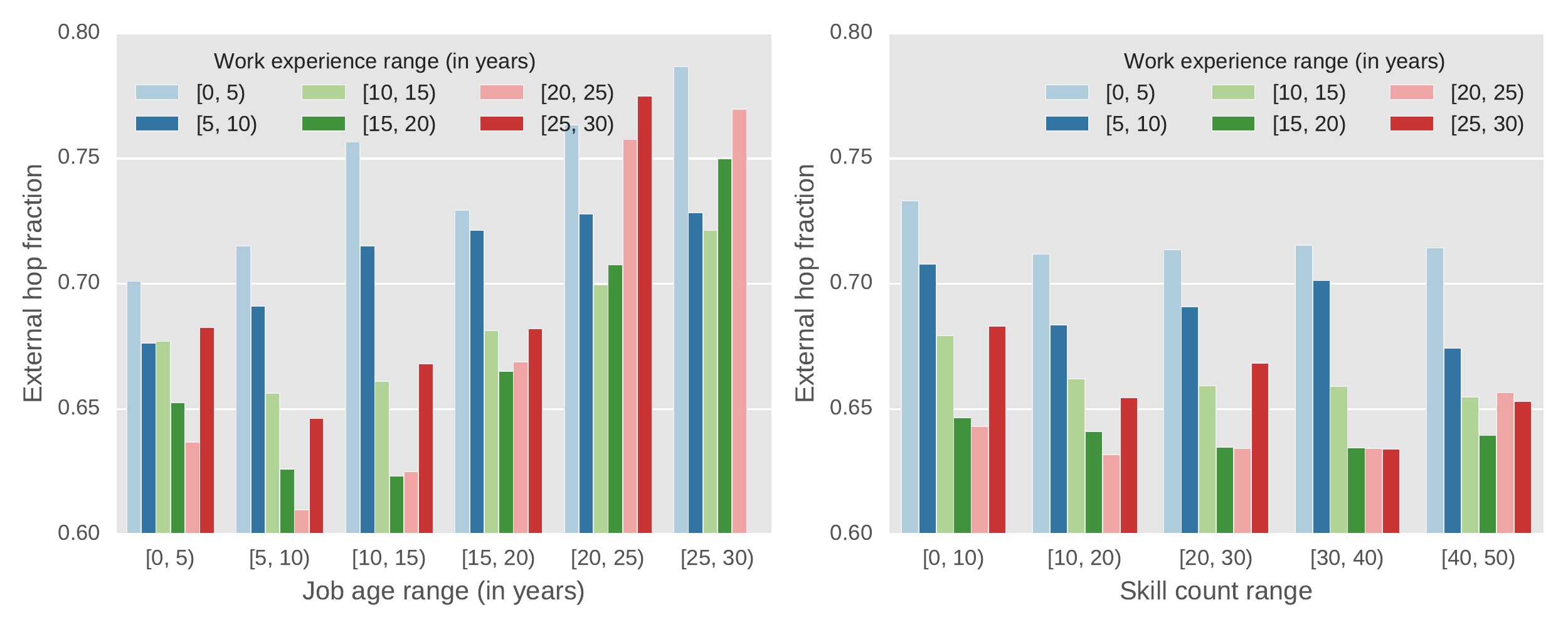}
\caption{Distributions over external hop fractions}
\label{fig:hop_frac_sliced_dist}
\end{figure*}

% \begin{figure}[!t]
% \centering
% \includegraphics[width=1.0\textwidth]{figures/hop_fraction_heatmap.pdf}
% \caption{Hop fraction heatmap}
% \label{fig:hop_frac_heatmap}
% \end{figure}

\subsection{External Hop Analysis}

It is also interesting to see how the external hop fraction (cf. Equation~\ref{eqn:ext_hop_frac}) varies with different combinations of job age, work experience, and skill count groups. Figure~\ref{fig:hop_frac_sliced_dist} summarizes the results, whereby we set the minimum support of $100$ for each bar in the plots. The left chart of Figure~\ref{fig:hop_frac_sliced_dist} reveals two insights:
\begin{itemize}
\item \textbf{External hop fraction vs users' work experience.} For all job age groups under $20$ years, younger workforce (with shorter work experience) is more likely to leave their jobs for other organizations than more experienced people. An exception here is people with $\geq 25$ years of work experience, whereby the external hop fraction shoots up. Again, these refer to very seasoned people (e.g., Director) whose skills are versatile and can freely hop to different organizations. For the job age groups of $\geq 20$ years, we can see more profound increase in the external hop fractions for people with work experience of $\geq 15$ years\footnote{We do not show the work experience range $[25,30)$ in the rightmost bar group of Figure~\ref{fig:hop_frac_sliced_dist} (left chart), since the number of support is less than $100$.}. These may reflect the tipping points for the seasoned workforce to find new job opportunities outside. Further investigation into this phenomenon is therefore an interesting future work.

\item \textbf{External hop fraction vs job age.} Comparisons can also be made among the external hop fractions for the same work experiences across different job age groups (i.e., bars of the same color). We can see that the external hop fraction tends to increase as the job age increases (i.e., more established jobs). This suggests that the older jobs are more likely to see competitions for human capital in general, and so it is more preferable to take up newer, trendier jobs. %An exception is seasoned workforce with $\geq 10$ years of work experience who take up new jobs with job age of $< 5$ years. This suggests that the younger workforce is more preferable than the older ones for new, emerging jobs.
\end{itemize}

\textbf{External hop fraction vs number of skills.}  Next, we wish to answer the question of whether more skills lead to higher external hops. The right chart of Figure~\ref{fig:hop_frac_sliced_dist} shows that, within each skill count group, younger people with less work experiences are more likely to move out to other organizations---except again for the very seasoned people with work experience of $\geq 25$ years. We note that this finding is consistent with that of several earlier works \cite{NBERw2649,Joseph:2012,Schawbel:2013}. Nonetheless, we do not see a clear association between the external hop fraction and the number of skills people have. That is, the propensity to hop out has little to do with the diversity of skills that people have.

\begin{table}[!t]
%\scriptsize
\centering
\caption{Count statistics of various hop types}
\label{tab:hop_type_comparisons}
\begin{tabular}{|l|c|c|c|}
\hline
     & Promotion & Demotion & Total \\
\hline
External hop & 4,813 & 1,720 & 6,533 \\
Internal hop & 3,891 & 387   & 4,278 \\
\hline
Total        & 8,704 & 2,107 & 10,811 \\
\hline
\end{tabular}
\end{table}

\begin{figure*}[!t]
\centering
\includegraphics[width=1.0\textwidth]{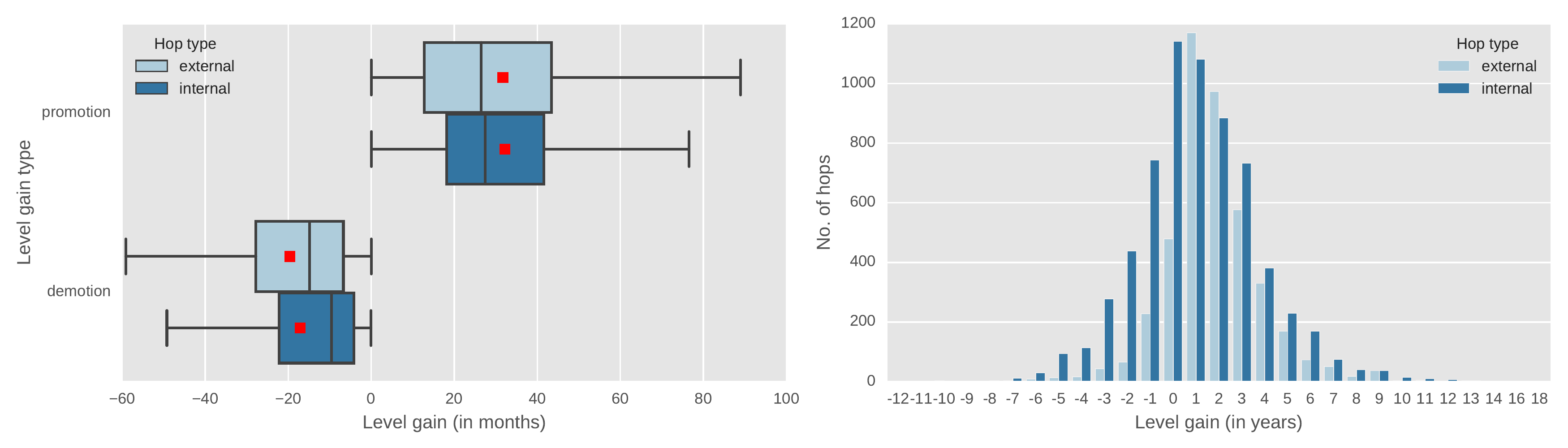}
\caption{Comparison of hop level gains for different hop types}
\label{fig:hop_type_comparisons}
\end{figure*}

\subsection{Promotion and Demotion Analysis}

As promotion is often a cited reason for people leaving one job for another.  We now conduct a promotion and demotion analysis by dividing the hops into external and internal hops based on level gain (i.e., promotion vs. demotion). Table~\ref{tab:hop_type_comparisons} and Figure~\ref{fig:hop_type_comparisons} summarize the results. To get a reliable estimate of level gain, and thus reliable ``promotion'' or ``demotion'' labels, we require both source and target jobs for each hop must fulfill the (default) minimum support of $10$. As such, we do not include in Table~\ref{tab:hop_type_comparisons} and Figure~\ref{fig:hop_type_comparisons} hops that fail to meet the minimum support.

From Table~\ref{tab:hop_type_comparisons}, we can derive two conclusions:
\begin{itemize}
  \item Firstly, the probability of promotion is greater than that of demotion, i.e., $p(\text{promotion}) = \frac{8,704}{10,811} = 80.51\%$ is greater than $p(\text{demotion}) = \frac{2,107}{10,811} = 19.49\%$.
  \item Secondly, people are more likely to get promoted due to internal hops than getting promoted due to external hops. That is, $p(\text{promotion} | \text{internal hop}) = \frac{3,891}{4,278} = 90.95\%$ is greater than $p(\text{promotion} | \text{external hop}) = \frac{4,813}{6,533} = 73.67\%$.
\end{itemize}
Figure~\ref{fig:hop_type_comparisons} shows a more fine-grained detail in terms of the level gain distribution. It is evident that the majority of the level gain values are positive, again suggesting that hopping most likely involves promotion rather than demotion (i.e., $p(\text{promotion}) > p(\text{demotion})$). However, we do not find clear differences between external and internal hops in terms of the level gain distribution. This observation is similar to the findings in \cite{Hamori:2010}.

\begin{figure*}[!t]
\centering
\includegraphics[width=0.8\textwidth]{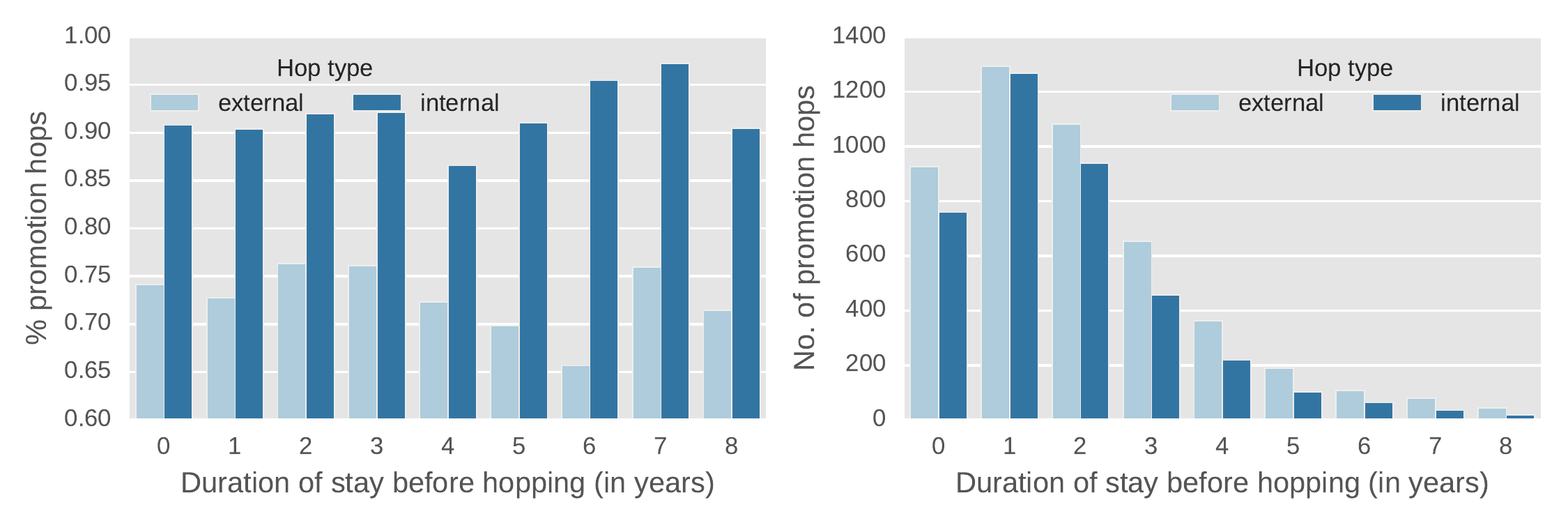}
\caption{Promotion hop fraction and counts for different durations of stay}
\label{fig:hop_vs_duration}
\end{figure*}

In addition, we investigate whether promotion hops vary with the duration of stay (at some job) before hopping. Figure~\ref{fig:hop_vs_duration} shows the promotion hop fractions (i.e., $p(\text{promotion} | \text{external hop})$ and $p(\text{promotion} | \text{internal hop})$) as well as promotion hop counts as a function of duration of stay prior to hopping. For these plots, we also set the minimum support threshold to filter out unreliable statistics. The right chart of Figure~\ref{fig:hop_vs_duration} suggests that promotion hops most commonly happen after a person works for $1$--$2$ years. However, the left chart of Figure~\ref{fig:hop_vs_duration} indicates no obvious relationship between the duration of stay and promotion hop fraction. Regardless, it is again evident that the probability of promotion is higher for internal hops than for external hops.

\subsection{Network Analysis}

\begin{table*}[!t]
%\scriptsize
\centering
\caption{Statistics of the job and organization graphs}
\label{tab:component_stats}
\begin{tabular}{|l|c|c|}
\hline
Metric       & Job graph & Organization graph \\
\hline
\multicolumn{3}{|c|}{Basic}\\
\hline
No. of nodes & 27,451 & 6,139  \\
No. of edges & 93,283 & 173,993 \\
Sparsity of adjacency matrix     & 0.01\% & 0.46\%\\
\hline
\multicolumn{3}{|c|}{Strongly connected component}\\
\hline
No. of SCCs                          & 15,455 			 & 415\\
Size of the largest SCC       		  & 11,950 (43.53\%) & 5,725 (93.26\%)\\
Size of the 2nd-largest SCC  	  & 4 (0.01\%)                  & 1 (0.02\%)\\
\hline
\multicolumn{3}{|c|}{Weakly connected component}\\
\hline
No. of WCCs                          & 882  			 & 15\\
Size of the largest WCC        	  & 25,747 (93.79\%)     & 6,125 (99.77\%)\\
Size of the 2nd-largest WCC          & 13 (0.05\%)      & 1 (0.02\%) \\
% \hline
% Clustering & Average no. of triangles 			  & 11.7572 & 770.0710 \\
% 		   & Transitivity 			 			  & 0.0441  & 0.1059 \\
% 		   & Average clustering coefficient 	  & 0.0003  & 0.0001 \\
\hline
\multicolumn{3}{l}{SCC = strongly-connected component, WCC = weakly-connected component}
\end{tabular}
\end{table*}

\textbf{Network structure analysis.}  In this section, we analyze the job hop behavior at the network level, which includes job and organization graphs (cf. Section~\ref{sec:hop_network}). The basic statistics of the job and organization graphs are summarized in Table~\ref{tab:component_stats}. We can conclude that both graphs are sparse in general, having small number of edges relative to the squared number of nodes. We also examine the connectedness of the graphs by looking at the strongly-connected component (SCC) and weakly-connected component (WCC) metrics. The former checks for connectedness by following the directionality of the graph edges, whereas the latter ignores the directionality. Overall, the results in Table~\ref{tab:component_stats} indicate that there exists a giant component for both job and organization graphs, and its size is significantly bigger than the second largest component. As such, we can conclude that our job and organization graphs are fairly well-connected, in the sense that there exists a path between any two nodes within the giant components.

% \begin{figure}[!t]
% \centering
% \includegraphics[width=1.0\textwidth]{figures/job_hop_component_dist.pdf}
% \caption{Distributions of strongly connected components in job hop graph}
% \label{fig:scc_dist_job}
% \end{figure}

% \subsubsection{Observations for Figure~\ref{fig:scc_dist_job}}

% \begin{itemize}
% \item There exists one giant connected component in the job hop graph, either strongly- or weakly-connected component. The other components are much smaller by comparison.
% \end{itemize}

% \begin{figure}[!t]
% \centering
% \includegraphics[width=1.0\textwidth]{figures/com_hop_component_dist.pdf}
% \caption{Distributions of strongly connected components in  hop graph}
% \label{fig:scc_dist_}
% \end{figure}

% \subsubsection{Observations for Figure~\ref{fig:scc_dist_}}

% \begin{itemize}
% \item There exists one giant connected component in the  hop graph, either strongly- or weakly-connected component. The other components are much smaller by comparison.
% \end{itemize}

\begin{figure*}[!t]
\centering
\includegraphics[width=0.8\textwidth]{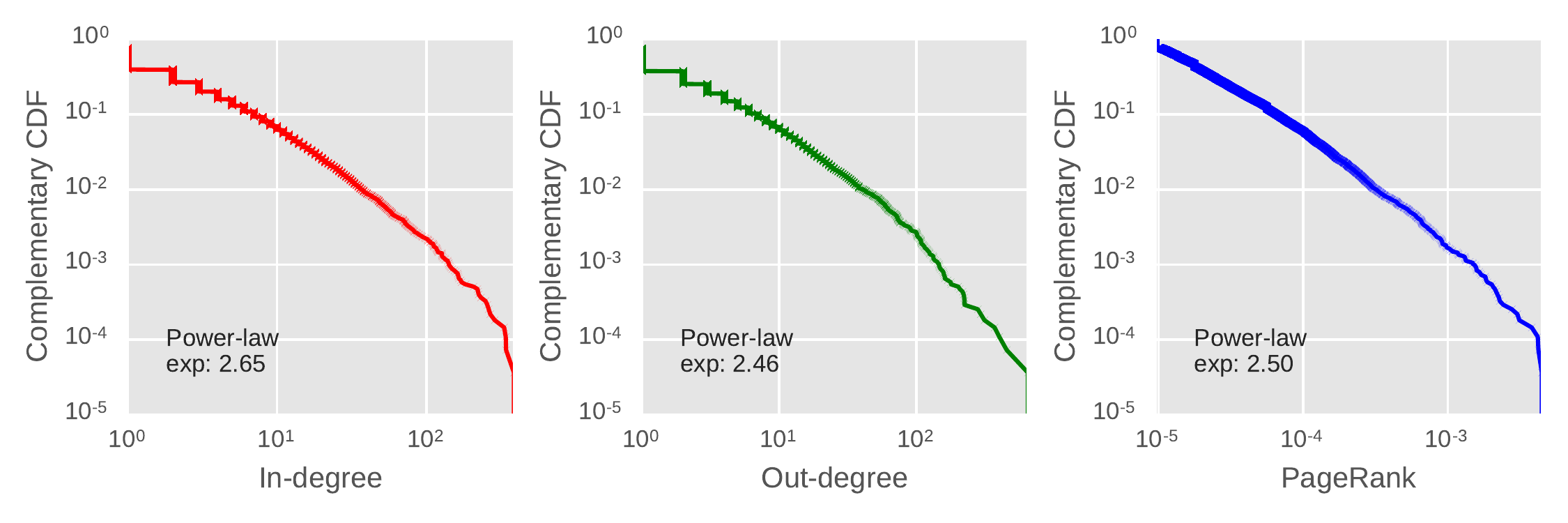}
\caption{Centrality distributions of job hop graph}
\label{fig:centrality_job}
\end{figure*}

% \begin{figure}[!t]
% \centering
% \includegraphics[width=1.0\textwidth]{figures/com_hop_centrality_dist.pdf}
% \caption{Centrality distributions of  hop graph}
% \label{fig:centrality_}
% \end{figure}

With the connectedness trait validated, we now examine the centrality properties of the nodes in our hop graphs. Figure~\ref{fig:centrality_job} presents the complementary cumulative distribution functions (CDFs) of the in-degree, out-degree, and PageRank centralities (cf. Section~\ref{sec:metrics}) for the job graph. It is shown that all three metrics exhibit heavy-tail, skewed distribution. We performed power-law fitting and obtained exponent terms of greater than $2$ for all graphs, thereby indicating a scale-free phenomenon. Similar result was obtained for the organization graph, although the results are not shown here due to space constraint.

\begin{figure*}[!t]
\centering
\includegraphics[width=0.85\textwidth]{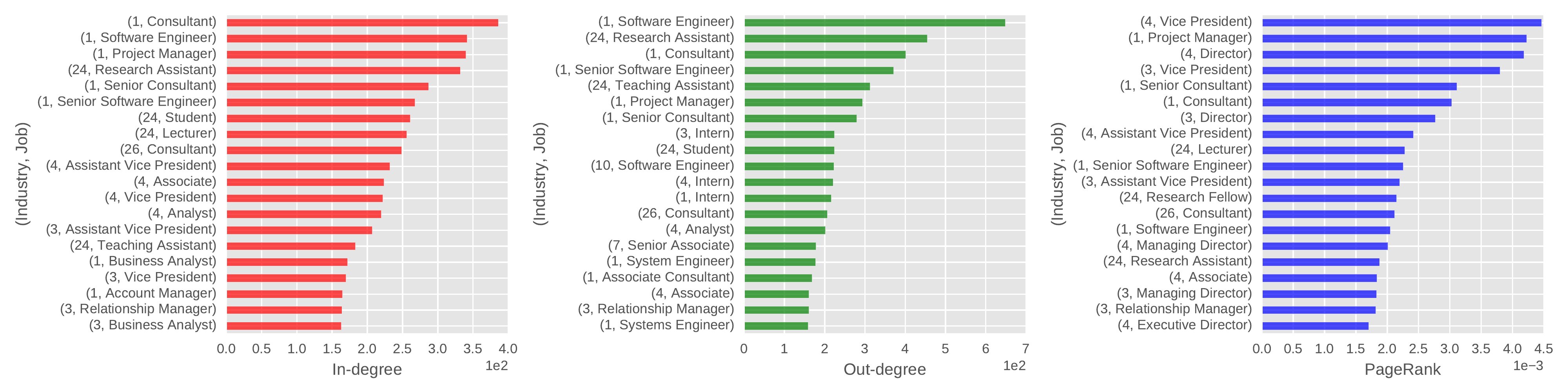}
\caption{Top 20 nodes in job graph with the largest centralities}
\label{fig:top_centrality_job}
\end{figure*}

\begin{figure*}[!t]
\centering
\includegraphics[width=0.85\textwidth]{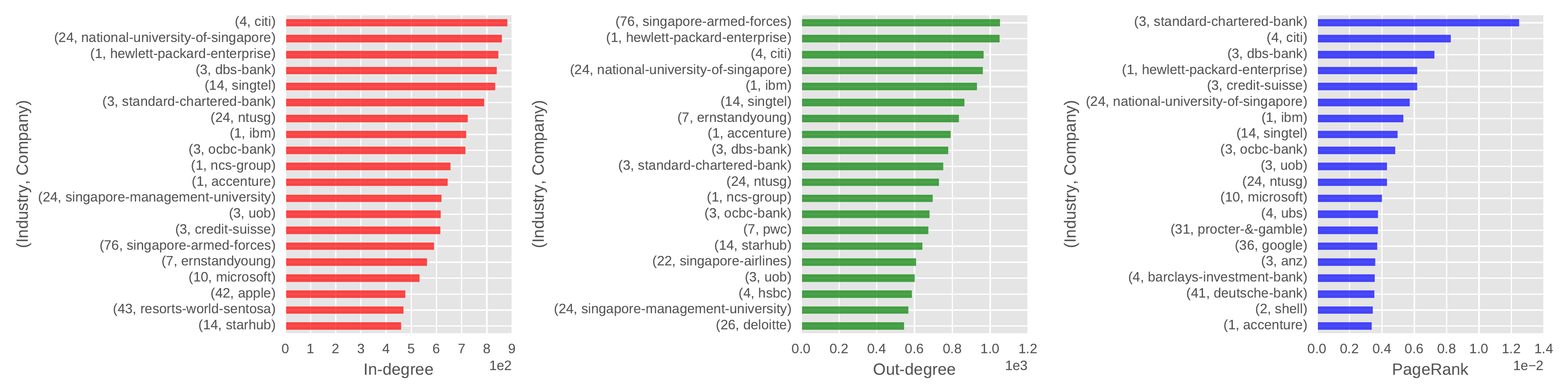}
\caption{Top 20 nodes in organization graph with the largest centralities}
\label{fig:top_centrality_}
\end{figure*}

\textbf{Job centrality analysis.} Next, we evaluate the top nodes having the highest centrality values in the job-level and organization-level graphs, as shown in Figures~\ref{fig:top_centrality_job} and \ref{fig:top_centrality_} respectively. The results provide several interesting insights. For the job graph, we find that the top in-degree, out-degree and PageRank jobs are overall dominated by major industries\footnote{Major industry codes: $1 =$ Information Technology and Services, $3 =$ Banking, $4 =$ Financial Services, $10 =$ Computer Software, $24 =$ Higher Education, $26 =$ Management Consulting}. From the left chart of Figure~\ref{fig:top_centrality_job}, we can see that the top in-degree nodes refer to those popular jobs in major industries that attract talents. Meanwhile, the middle chart of Figure~\ref{fig:top_centrality_job} suggests that the top out-degree jobs are those that involve versatile skills (e.g., software engineer, consultant) or interim roles (e.g., intern). People having these jobs may thus be able to move to more diverse range of jobs/organizations (i.e., talent supplier). Finally, the right chart of Figure~\ref{fig:top_centrality_job} reveals that the top PageRank nodes refer to high-level, managerial jobs (e.g., Director, Manager, Vice President). This conforms with our intuition about PageRank as a measure of the desirability of a job (cf. Section \ref{sec:metrics}).

\section{Conclusion}
\label{sec:conclusion}

In this paper, we put forward a data analytics approach to study job hops at a large scale using data from a city state's OPN. Our study leads to a few key  takeaways. Firstly, from our distribution analyses, we discovered that: (1) younger workforce with less work experience is more likely to move out to other companies than more experienced ones, except for very seasoned people; (2) the propensity to switch jobs to other organizations is higher for older jobs, which are more likely to see competitions for human capital than the newer ones; and (3) job hops involve promotions more likely than demotions, and people are more likely to get promoted due to internal hops than getting promoted due to external hops;
%and (4) hops that involve promotion do not depend on the duration of stay prior to hopping.
Secondly, from our hop network analyses, we found that: (1) top in-degree job (organization) nodes are prominent jobs (companies) that attract talents, whereas top out-degree job (organization) nodes are influential jobs (organizations) that  supply talents; and (2) job (organization) nodes with high PageRank refer to desirable, major jobs (organizations) that are well-known for providing good career offering.
% and (3) With respect to each company in the organization-level graph, promotion due to internal hops is more likely than promotion due to external hops

The findings from this paper lead to a few possibilities.  Firstly, we demonstrate that it is possible to repurpose the career histories of OPN profiles to study the job hop patterns of workforce within a city or country. This vastly improves the scale and granularity of job hop study which was traditionally done using surveys.  Through our analysis, we show that the propensity to perform job hops is relatively higher among the young workforce than the older one.  This could lead to two main concerns, namely: (i) the limited time to acquire adequate skills on the job among the young employees; and (ii) the unwillingness of companies to provide them skill training.  These concerns may cost the workforce long-term's skill development and productivity. To overcome these concerns, more incentives may be introduced to encourage young employees to stay longer on their jobs.  One could also increase the chance of job promotions among the younger employees.

Our analysis also shows that job and organization graphs are well connected. We further define job centrality measures to determine attractive jobs and companies.  Such measures allow jobs and companies to be ranked for applicants' reference during job search.  These measures can also further refined to find attractive jobs and companies in specific industry domains. 

% In this paper, we present a data-driven approach to profile job hops at a large scale using public user profiles in OPN. We introduced the work experience and job age metrics to measure the seniority and establishment of a job, respectively, and then analyze on how they relate to the propensity of hopping. We also studied how the job hop is related to promotion/demotion of employees, and compare the probability of promotion for external and internal hops. Lastly, we investigated on the centralities of the job-level and organization-level hop graphs in order to understand talent flow and job/organizational competitiveness. Moving forward, we would like to extend our hop analytics approach to develop a new career path optimization and recommendation methodology. 

\section*{Acknowledgement}
This research is supported by the Singapore National Research Foundation under its International Research Centre@Singapore Funding Initiative and administered by the IDM Programme Office, Media Development Authority (MDA).

% number - used to balance the columns on the last page
% adjust value as needed - may need to be readjusted if
% the document is modified later
%\IEEEtriggeratref{8}
% The "triggered" command can be changed if desired:
%\IEEEtriggercmd{\enlargethispage{-5in}}

% references section

% can use a bibliography generated by BibTeX as a .bbl file
% BibTeX documentation can be easily obtained at:
% http://mirror.ctan.org/biblio/bibtex/contrib/doc/
% The IEEEtran BibTeX style support page is at:
% http://www.michaelshell.org/tex/ieeetran/bibtex/
%\bibliographystyle{IEEEtran}
% argument is your BibTeX string definitions and bibliography database(s)
%\bibliography{IEEEabrv,../bib/paper}
%
% <OR> manually copy in the resultant .bbl file
% set second argument of \begin to the number of references
% (used to reserve space for the reference number labels box)

\bibliographystyle{IEEEtran}
\bibliography{references}

% that's all folks
\end{document}